\documentclass[twocolumn,showpacs]{revtex4}
\usepackage{amsmath,amssymb,dcolumn}
\usepackage{epsfig}
\def\imo{i}
\begin{document}
\title{Stability of higher dimensional Reissner-Nordstr\"om-anti-de Sitter black holes}
\author{R. A. Konoplya}\email{konoplya_roma@yahoo.com}
\affiliation{Department of Physics, Kyoto University, Kyoto 606-8501, Japan}
\author{A. Zhidenko}\email{zhidenko@fma.if.usp.br}
\affiliation{Instituto de F\'{\i}sica, Universidade de S\~{a}o Paulo \\
C.P. 66318, 05315-970, S\~{a}o Paulo-SP, Brazil}
\begin{abstract}
We investigate stability of the D-dimensional Reissner-Nordstr\"om-anti-de-Sitter metrics  as solutions of the Einstein-Maxwell equations. We have shown that asymptotically anti-de Sitter black holes are dynamically stable for all values of charge and anti-de Sitter radius in $D=5,6\ldots11$ dimensional space-times. This does not contradict to dynamical instability of RN-AdS black holes found by Gubser in $\mathcal{N}=8$ gauged supergravity, because the latter instability comes from the tachyon mode of the scalar field, coupled to the system.
Asymptotically AdS black holes are known to be thermodynamically unstable for some region of parameters, yet, as we have shown here, they are stable against gravitational perturbations.

\end{abstract}
\maketitle
\section{Introduction}

Nowadays, brane world theories and  string theory, imply existence of extra dimensions in nature \cite{brane_string}. This induced great interest to higher dimensional black holes, in particular, to such features as perturbations, dynamical and thermodynamical stability, particles and fields behavior and Hawking radiation around these black holes. Unlike four dimensional case, higher dimensional space-times, admit plenty of ``black'' solutions: black holes, black strings and branes, black rings and Saturns. Stability of these solutions may be criteria of their existence. The stability analysis usually requires to transform the perturbed Einstein equation to the wave-like form, what was done yet in 1957 by Regge and Wheeler for $D=4$ black holes, and only in 2003 for general number of space-time dimensions \cite{ishibashi_kodama}. Then, the stability of D-dimensional Schwarzschild black holes was proved in  \cite{ishibashi_kodama2} and of Schwarzschild-de Sitter black holes in \cite{Konoplya:2007jv}. Recently the stability of Kaluza-Klein black holes with squashed horizons was shown in \cite{Ishihara:2008re}, \cite{Kimura:2007cr}.
Unlike Kaluza-Klein black holes, black strings and branes become unstable for perturbations with wavelength, which is larger than some threshold value   (Gregory-Laflamme instability \cite{Gregory:1993vy}). In  \cite{Konoplya:2008yy} it was shown  that the instability  threshold point corresponds to some dominating static solution of the wave equation. The neutral D-dimensional black holes in Gauss-Bonnet theory are unstable only for $D=5, 6$ and for small values of Gauss-Bonnet coupling \cite{Konoplya:2008ix},  \cite{Konoplya:2008ix2}. The instability in the Gauss-Bonnet theory is qualitatively different from a black string instability: the black string instability is an example of instability, developed at the lowest multipoles, therefore with a static solution dominance at the threshold point. The instability in the Gauss-Bonnet theory is developed at large multipoles, so that the growing mode dominates after a long period of damped quasinormal oscillations \cite{Konoplya:2008ix2}.

The stability of the higher-dimensional black holes is important also for the growing interest to the quasinormal modes of the Standard Model fields
in higher dimensional theories \cite{highD_qnm}. Indeed, only stable black holes can be considered as a background on which, test fields propagate.

Higher dimensional black holes in asymptotically anti-de Sitter (AdS) space-times have been in the focus of string theorists recent years, because of their role in the AdS/CFT correspondence. A large asymptotically anti-de Sitter black hole corresponds to a thermal state in the dual conformal field theory, where the Hawking temperature of the black hole is the temperature in the dual field theory \cite{Witten:1998zw}. The perturbations of AdS black holes have been extensively studied during recent decade \cite{AdS_perturbations}. Nevertheless, it was not known until the present study, if D-dimensional asymptotically AdS black holes are dynamically stable as solutions of D-dimensional Einstein-Maxwell equations. The stability of $D=4$ and $D=5$ Reissner-Nordstr\"om-AdS black holes were studied by Gubser and Mitra  in the
$\mathcal{N}=8$ gauged super-gravity theory \cite{Gubser:2000ec} \cite{Gubser:2000mm}, i .e. a theory with the Maxwell and scalar matter fields coupled to the electromagnetic field. There it was shown that the highly charged black holes
are unstable and the parameter region of instability increases for larger black holes, i.e. for black holes,
which radius is much larger than the anti-de Sitter radius. Yet that instability evidently came from the tachyonic mode of the scalar field.
Thus the question remains if there is an instability of Reissner-Nordstr\"om-anti-de Sitter black holes within the ordinary Einstein-Maxwell theory?
Our main aim here is to answer this question, keeping in mind such an important feature of black holes as thermodynamic (in)stability.
Thus, according to hypothesis of Gubser and Mitra  in \cite{Gubser:2000ec}  and \cite{Gubser:2000mm}, there may be a correlation between thermodynamic and dynamic (gravitational) (in)stabilities of black holes, because it was found, that the parametric region of thermodynamic and dynamic (in)stabilities although do not coincide, differ from each other only slightly for the $\mathcal{N}=8$ gauged super-gravity.
Another possible correlation could give the thermodynamic instability of small AdS black holes, what happens with a phase transition, called the Hawking-Page transition \cite{Hawking-Page}. One of the thermodynamically preferred final states, after the transition, might be a pure AdS space-time \cite{Hawking-Page}.

The thermodynamic instability of black holes takes place also for ordinary Einstein-Maxwell-AdS black holes \cite{Chamblin:1999hg} for some values of black hole parameters. The second order phase transition occurs at the instability point \cite{Chamblin:1999hg}.
If we expect some correlations between thermodynamic and dynamic instabilities to these cases, we should expect gravitational instability of D-dimensional charged black holes in AdS space-times within the standard Einstein-Maxwell theory. In this paper, we shall show that this is not the case of the pure RNAdS black holes, which are stable against gravitational perturbations for $D = 5,6\ldots11$, where $D$ is the number of space-time dimensions.


The paper is organized as follows: Section II introduces the basic formula for the background metric and for perturbation equations reduced to a wave-like form. Section III reviews the numerical method, which we used for stability analysis. Sections IV and V consider the obtained results for Reissner-Nordstr\"om -anti- de Sitter black holes.

\section{Basic formulae}

The metric of the $D=d+2$-dimensional Reissner-Nordstr\"om-(anti)-de-Sitter black
holes is given by the line element
\begin{equation}\label{metric}
ds^2=f(r)dt^2-\frac{dr^2}{f(r)}-r^2d\Omega_d,
\end{equation}
where $d\Omega_d$ is the line element on a unit $d$-sphere, and
\begin{equation}\label{metric-function}
f(r)=1-X+Z-Y,
\end{equation}
$$X=\frac{2M}{r^{d-1}},\qquad Y=\frac{2\Lambda r^2}{d(d+1)}, \qquad Z=\frac{Q^2}{r^{2d-2}}.$$

\begin{figure}
\includegraphics[width= 0.7 \linewidth]{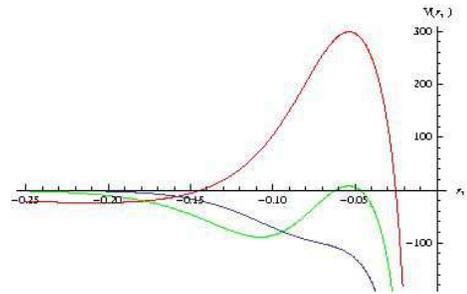}
\caption{Effective potential for scalar type of gravitational perturbations of the RNAdS black holes for $D=5$, $r_+=10 R$.}\label{D=5.r=10potential}
\end{figure}

Here $M$ is the mass parameter of the black hole, $Q$ is its charge, and the $\Lambda$-term coincides with a cosmological constant,
when positive, and is related to the anti-de Sitter radius, when negative, in the following way,

 $$\frac{2\Lambda}{d(d+1)}=-\frac{1}{R^2}=-1.$$

The general perturbations of the Einstein-Maxwell equations
\begin{equation}
g_{\mu \nu} =  g_{\mu \nu}^{0} + \delta g_{\mu \nu},
\end{equation}
\begin{equation}
\delta R_{\mu \nu} = \kappa~\delta\left( T_{\mu \nu} -\frac{1}{D-2}T g_{\mu\nu}\right)+\frac{2\Lambda}{D-2}\delta g_{\mu\nu},
\end{equation}
after using the gauge freedom, and separating the angular variables
can be reduced to a number of wave like equations  \cite{ishibashi_kodama} for three types of gravitational perturbations, according to the symmetry of the rotation group:  scalar, vector and tensor. In four dimensions, scalar type of gravitational perturbations is called polar,
vector type is called axial. Tensor type of gravitational perturbations is usually pure gauge in $D=4$ black hole space-times.
For $D=4$ Reissner-Nordstr\"om-(anti)-de Sitter black holes, axial and polar types of perturbations are isospectral, so that for stability analysis, it is enough to analyze only one type of perturbations. For $D > 4$, the isospectrality is broken, so that one needs to check all three kind of perturbations.  The vector and tensor types of gravitational perturbations were shown to be stable \cite{ishibashi_kodama} for all $D$, with the help of the so-called S-deformation technique. Therefore we shall consider here only the scalar type of gravitational perturbations.

The equation of motion for gravitational perturbations of scalar type can be reduced to the wave-like equation \cite{ishibashi_kodama2},
\begin{equation}\label{wave-like}
\left(\frac{d^2}{dr_*^2}+\omega^2-V_\pm\right)\Psi(r)=0,
\end{equation}
where \emph{the tortoise coordinate} $r_*$ is defined as
\begin{equation}\label{tortoise}
dr_*=\frac{dr}{f(r)},
\end{equation}
\begin{equation}\label{potential}
V_\pm(r)=f(r)\frac{U_\pm}{64r^2H_\pm^2}.
\end{equation}

Here $V_{+}$ and $V_{-}$ are potentials for the two kinds of scalar gravitational perturbations. The potential $V_{-}$ reduces to the pure gravitational perturbations when the black hole charge vanishes, while $V_{+}$ reduces to the perturbations
of the test Maxwell field in the black hole background in this limit. When the charge $Q$ is non-zero, the gravitational and electromagnetic perturbations are coupled.

Note, that $V_+$ is proven to be stable \cite{ishibashi_kodama} with the help of the S-deformation. Thus we are left with the $V_-$ potential, which must be tested on stability.

Here we used the values

\begin{widetext}
\begin{eqnarray}
&& H_-=\lambda+\frac{d(d+1)}{2}(1+\lambda\delta)X,
\end{eqnarray}
\begin{eqnarray}
%
& U_- =
  & \left[-4 d^3 (d+2) (d+1)^2 (1+\lambda \delta)^2 X^2
      +48 d^2 (d+1) (d-2) \lambda (1+\lambda \delta) X  \right.
\notag\\
&& \left.  -16 (d-2) (d-4) \lambda^2\right] Y
     -d^3 (3 d-2) (d+1)^4 \delta (1+\lambda \delta)^3 X^4
\notag\\
&& -4 d^2 (d+1)^2 (1+\lambda \delta)^2
     \left\{(d+1)(3 d-2) \lambda \delta-d^2\right\} X^3
\notag\\
&&  +4 (d+1) (1+\lambda \delta)\left\{ \lambda (d-2) (d-4) (d+1) (\lambda+d^2 ) \delta
  \right. \notag\\
&& \left. \quad  +4 d (2 d^2-3 d+4) \lambda+d^2 (d-2) (d-4) (d+1) \right\}X^2
\notag\\
&&  -16\lambda \left\{ (d+1) \lambda \left(-4 \lambda+3 d^2(d-2) \right) \delta
\right.\notag\\
&&\left.  +3 d (d-4) \lambda+3 d^2 (d+1) (d-2) \right\}X
\notag\\
&&      +64 \lambda^3+16 d(d+2)\lambda^2 .
\end{eqnarray}
\end{widetext}

We shall imply that $$\Psi \sim e^{-i \omega t}, \quad \omega = \omega_{Re} - i \omega_{Im},$$ so that
$\omega_{Im} > 0$ corresponds to a stable (decayed) mode, while $\omega_{Im} < 0$ corresponds to an unstable (growing) mode.
If the effective potential
$V(r)$ is positive definite everywhere outside the black hole event horizon, the differential operator
$$\frac{d^{2}}{dr_{*}^{2}} + \omega^{2}$$
is positive self-adjoint operator in the Hilbert space of the
square integrable functions of $r^{*}$, and, in that case any solution
of the wave equation with compact support is bounded, what implies
stability. An important feature of the gravitational perturbations is that
the effective potential $V_{-}$ (Eq.\ref{potential}), which governs the scalar type of the
perturbations, has negative gap for the higher dimensional black holes.
Therefore the instability is not excluded for this case, and numerical analysis of perturbations is necessary.

The values
$$2\lambda\delta=\sqrt{1+\frac{4\lambda Q^2}{(d+1)^2M^2}} -1,$$
$$\qquad\lambda=(\ell+d)(\ell-1), \quad\ell=2,3,4\ldots$$
are constants.

For convenience we shall parameterize the black hole mass and charge by its event horizon $r_+$ and inner horizon $r_-<r_+$ respectively. The value $r_-=0$ corresponds to the uncharged black hole.

\section{Numerical method}

For analysis of stability we need to test a black hole response to external perturbations, which is dominated by the so-called quasinormal modes at late time. The quasi-normal boundary conditions correspond to the pure out-going waves at infinity and pure in-coming waves
at the event (or de Sitter) horizon for asymptotically flat or de Sitter black holes. For asymptotically AdS black holes,
the Dirichlet boundary conditions are imposed at infinity. If growing modes exist, the considered system is unstable. Although usually,
damped quasinormal modes have both real and imaginary parts, i.e. are oscillating, the growing modes \cite{Konoplya:2008yy} are
\emph{non-oscillating}, that is \emph{pure imaginary}. This makes our search of unstable modes much easier.


Below we shall discuss the numerical method, which we used here for asymptotically anti-de Sitter space-times.

\begin{figure}
\includegraphics[width=0.7 \linewidth]{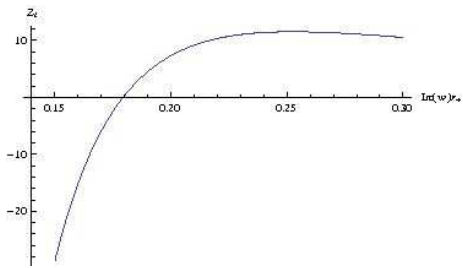}
\caption{Sample of instability for the Gauss-Bonnet black holes, $D=5$, $\ell=2$, $\alpha=0.5$.
Instability corresponds to zero of $Z_i$.}\label{GB}
\end{figure}

Let us start from the analysis of singularities of the equation (\ref{wave-like}). At the event horizon $V_\pm(r)\propto f(r)\sim 0$. Therefore
$$\Psi(r)\sim(r-r_+)^{\pm\imo\omega/f'(r_+)}.$$
The quasinormal boundary conditions at the event horizon imply

$$\Psi(r)=(r-r_+)^{-\imo\omega/f'(r_+)}(Z_0+{\cal O}(r-r_+)).$$

At the spatial infinity the two linear independent solutions of $\Psi(r)$ are
\begin{eqnarray}
\Psi_1(r)\sim r^{-(D-4)/2}, &\Psi_2(r)\sim r^{(D-6)/2}, & D\neq5,\label{gen-infinity}\\\nonumber
\Psi_1(r)\sim r^{-1/2}, & \Psi_2(r)\sim r^{-1/2}\ln(r), & D=5,
\end{eqnarray}

The quasi-normal boundary conditions imply that
\begin{equation}\label{infinity-BC}
\Psi(r\rightarrow\infty)\propto\left\{
                           \begin{array}{ll}
                             r^{-1}, & D=4; \\
                             r^{-1/2}, & D=5; \\
                             r^{-(D-4)/2}, & D\geq6.
                           \end{array}
                         \right.
\end{equation}

Let us consider the new function
\begin{equation}
y(r)=\left(\frac{r-r_+}{r-r_-}\right)^{\imo\omega/f'(r_+)}\Psi(r).
\end{equation}

If $\Psi(r)$ satisfies the quasi-normal boundary conditions, $y(r)$ is regular at the event horizon. Since the function $y(r)$ satisfies the linear equation, we fix its scale as
$$y(r_+)=1.$$
Then $y'(r_+)$ can be found from the equation (\ref{wave-like}),
$$y'(r_+)=\frac{\imo\omega f''(r_+)}{2f'(r_+)^2}-\frac{\imo\omega}{(r_+-r_-)f'(r_+)}+\frac{V_0}{f'(r_+)-2\imo\omega}\,,$$
where
$$V_0=\lim_{r\rightarrow r_+}\frac{V_\pm(r)}{f(r)}=\frac{U_\pm(r_+)}{64r_+^2H_\pm^2(r_+)}.$$

Imposing the above discussed boundary conditions at the event horizon, we solve the equation (\ref{wave-like}) numerically for each $\omega$ using the $NDSolve$ built-in function in \emph{Mathematica} for $r \geq r_f$, where $r_f\gg r_+$.

In the general case the behavior of $\Psi(r)$ at infinity is a superposition of the two solutions (\ref{gen-infinity}) for in-going and out going waves $\Phi_i(r)$, $\Phi_o(r)$,
\begin{equation}\label{fit-function}
\Psi_f(r)=Z_i \Phi_i(r)+Z_o\Phi_o(r),
\end{equation}
where $\Psi_o(r)$ satisfies the quasi-normal boundary condition (\ref{infinity-BC}). If $\omega$ is the quasi-normal frequency, the corresponding solution must satisfy the boundary conditions (\ref{infinity-BC}) at the spatial infinity and, thereby, $Z_i=0$.

Thus, our numerical procedure is the following. We integrate the equation (\ref{wave-like}) numerically imposing quasi-normal boundary condition at the event horizon. At large distance we compare the obtained function $\Psi(r_f)$ with (\ref{fit-function}) and find, thereby, the coefficients $Z_i$ and $Z_o$ for any given value of $\omega$. The quasi-normal modes correspond to the roots of the equation
\begin{equation}\label{QNM-equation}
Z_i(\omega)=0.
\end{equation}


In order to find $Z_i$ and $Z_o$ one has to find analytically expansions of $\Phi_i(r)$ and $\Phi_o(r)$ at large distance. The expansion
$$ \Phi_o= $$
\begin{equation}\label{serieo}\left\{
           \begin{array}{ll}
             r^{-1}, & D=4 \\
             r^{-\frac{D-4}{2}}, & D\geq5
           \end{array}
         \right\}\left(1\!+\!\frac{C^{(o)}_1}{r}\!+\!\frac{C^{(o)}_2}{r^2}\!+\!\frac{C^{(o)}_3}{r^3}\!\ldots\right)
\end{equation}
contains only inverse powers of $r$, while the expansion
$$ \Phi_i= $$
\begin{equation}\label{seriei}\left\{
           \begin{array}{ll}
             1, & D=4 \\
             r^{-1/2}\ln(r), & D=5 \\
             r^{\frac{D-6}{2}}, & D\geq6
           \end{array}\right\}\left(1+\frac{C^{(i)}_1}{r}+o\left(\frac{1}{r}\right)\right)
\end{equation}
contains also subdominant terms of the form of order $\displaystyle\frac{\ln(r)}{r}$. Since the series in (\ref{seriei}) are convergent we have used only the first term of the expansion, which does not contain logarithm. The expansion of (\ref{serieo}) was done up to the order $\sim r^{-3}$.

The analytical expansion allows to find $\Phi_i(r)$ and $\Phi_o(r)$ within the desired precision for $r\gg r_+$. If $\Psi(r_f)$ were known exactly, one would had found the coefficients $Z_i$ and $Z_o$ from the system of the linear equations
\begin{eqnarray}
\Psi(r_f)&=&\Psi_f(r_f),\\
\Psi'(r_f)&=&\Psi_f'(r_f).
\end{eqnarray}
In practice, being the result of the numerical integration, the values $\Psi(r_f)$ and $\Psi'(r_f)$ contain a numerical error, which causes low precision of the coefficients $Z_i$ and $Z_o$, found in this way. In order to minimize the numerical error, we find numerically the values of $\Psi$ at some large number of points near $r=r_f$. Then we fit the obtained numerical values of $\Psi$ by the function $\Psi_f(r)$ (\ref{fit-function}).
From the fit data we find $Z_i$ and $Z_o$ by solving the least squares problem at those points.

Since unstable modes are purely imaginary, one can restrict the searching area for $\omega$ by the imaginary axis. In this case the problem simplifies because the eigenfrequencies and the coefficients $Z_i$, $Z_o$ are real. It turns out, that the coefficient $Z_i$ changes its sign when crossing the solution. This can be used as an indicator of the existence of an unstable mode in the spectrum.

In order to be sure that the above method indeed can find an instability we tested it for the two cases when the instability is determined both analytically and numerically by alternative method. Namely, we checked the instability of the black strings \cite{Gregory:1993vy}, \cite{Konoplya:2008yy}, and also found the unstable modes of the Gauss-Bonnet black hole. Their values are in agreement with those, obtain within time-domain integration. Thus, as an example, on Fig. \ref{GB} one can see that the unstable mode is $\omega = 0.18 i$, what perfectly agrees with the value of unstable mode, obtained by the time-domain integration method in \cite{Konoplya:2008ix2}.

Another restriction upon the possible values of unstable modes comes from the depth of the negative potential gap ($V - \omega^2  > 0$ guarantees stability), $$Im(\omega)<\sqrt{-V_{min}},$$ where $V_{min}$ is the minimal value of the effective potential at $r_+\leq r<\infty$.

For complex quasi-normal frequencies $Z_i$ is complex. For this case the solution of (\ref{QNM-equation}) can be found by minimizing $|Z_i(\omega)|$. Unfortunately, due to oscillation of the solution in the asymptotically flat and asymptotically de Sitter backgrounds, we were unable to fit the solution at very large distances. However, for asymptotically AdS background the solution does not oscillate at large distance, and the described approach can be used to find quasi-normal modes of stable solutions.

\begin{figure}
\includegraphics[width= 0.7 \linewidth]{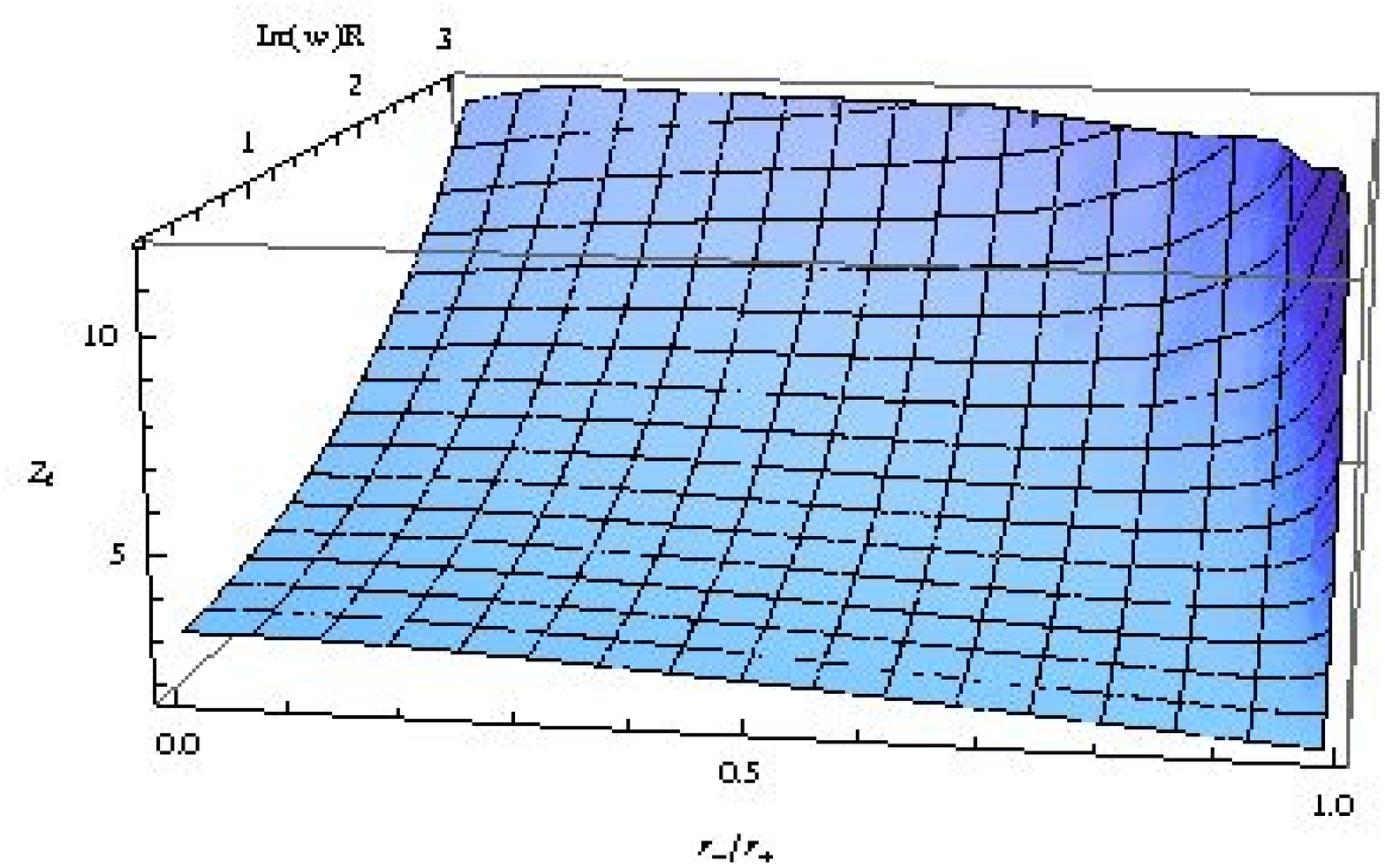}
\caption{$Z_i$ as a function of $r_-$ and $Im(\omega)$ for $D=5$, $r_+=1 R$}\label{D=5.r=1}
\end{figure}
\begin{figure}
\includegraphics[width= 0.7 \linewidth]{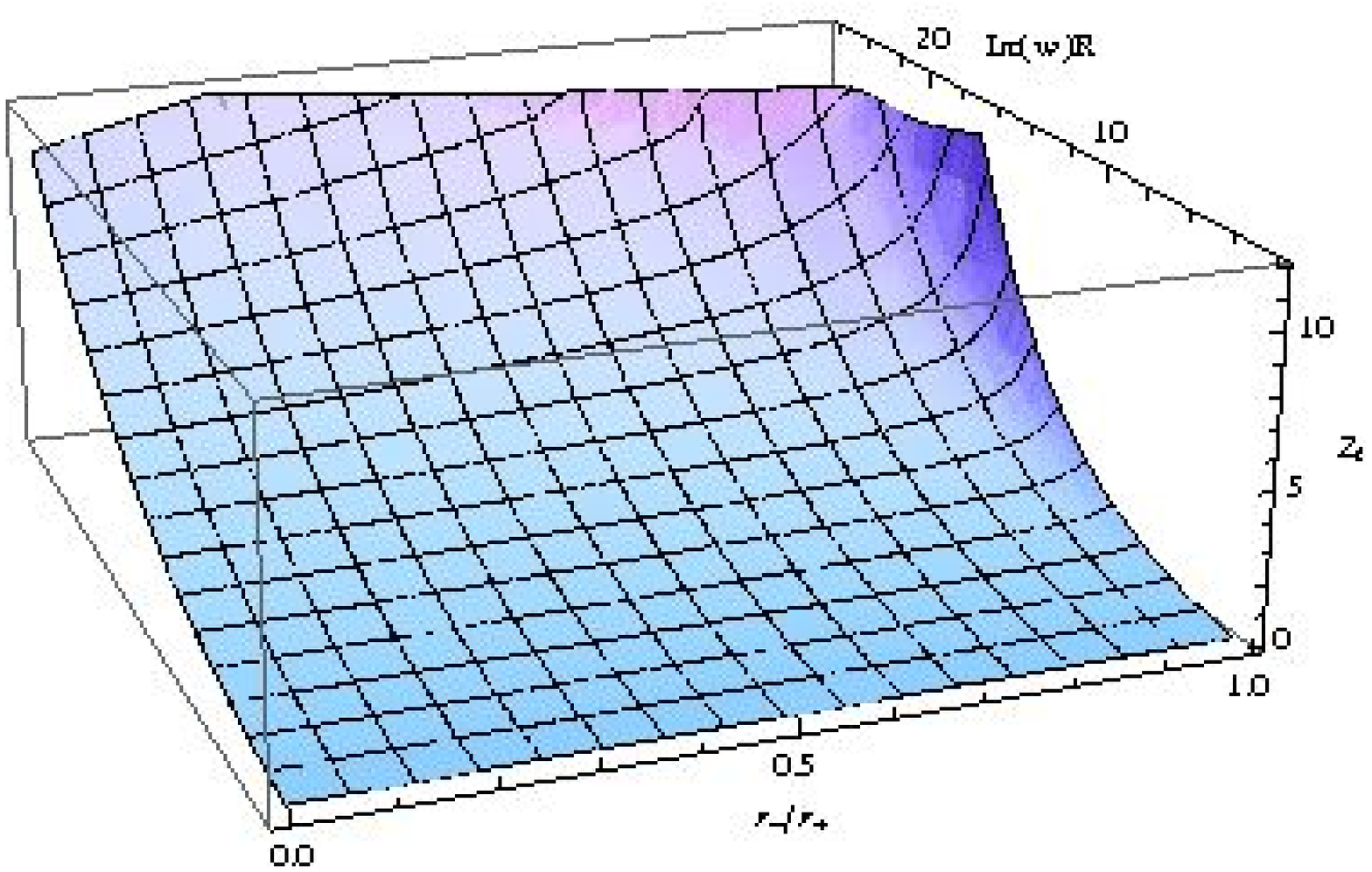}
\caption{$Z_i$ as a function of $r_-$ and $Im(\omega)$ for $D=5$, $r_+=10 R$}\label{D=5.r=10}
\end{figure}
\begin{figure}
\includegraphics[width=0.7 \linewidth]{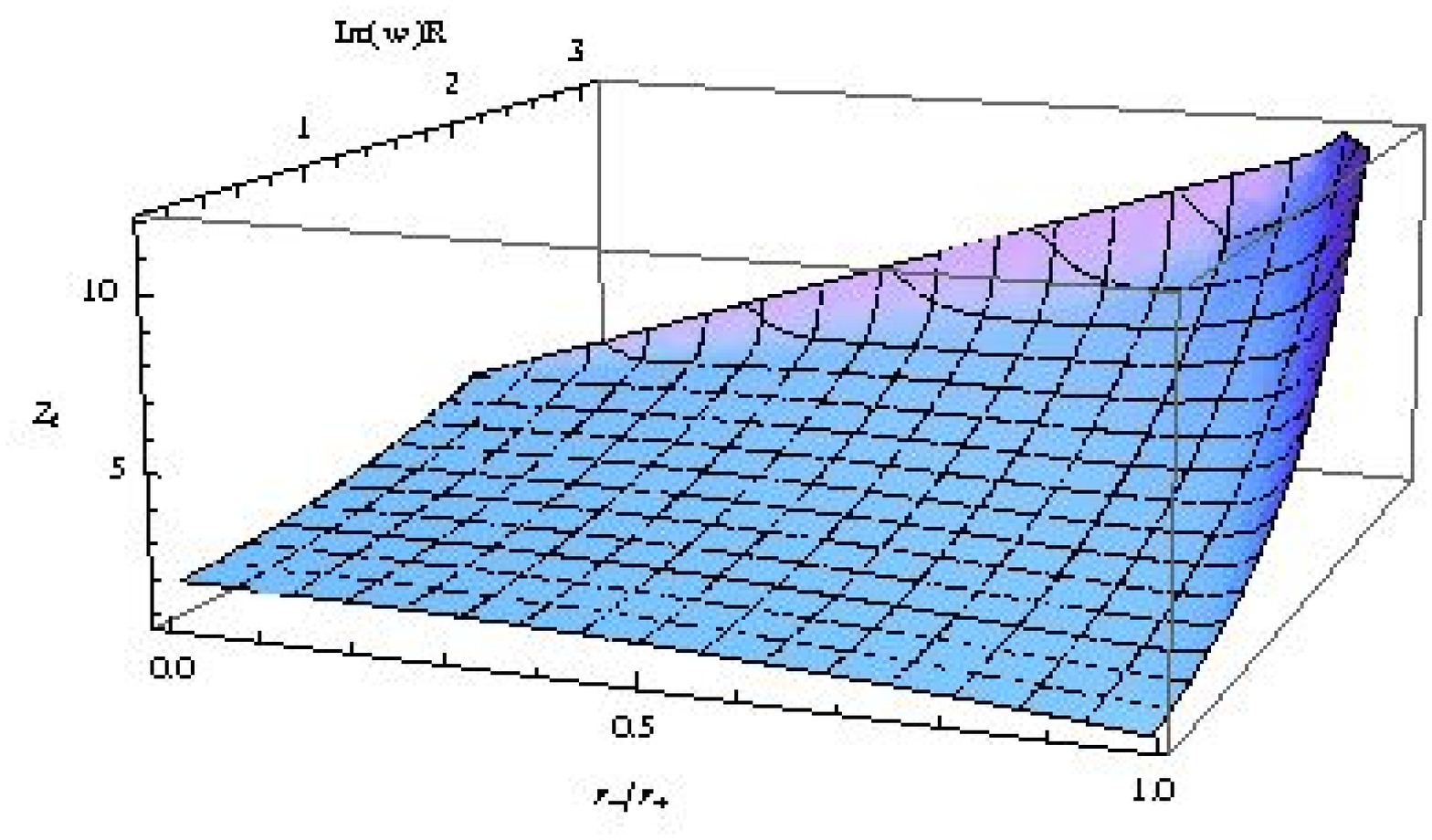}
\caption{$Z_i$ as a function of $r_-$ and $Im(\omega)$ for $D=6$, $r_+=R$.}\label{D=6.r=1}
\end{figure}
\begin{figure}
\includegraphics[width=0.7 \linewidth]{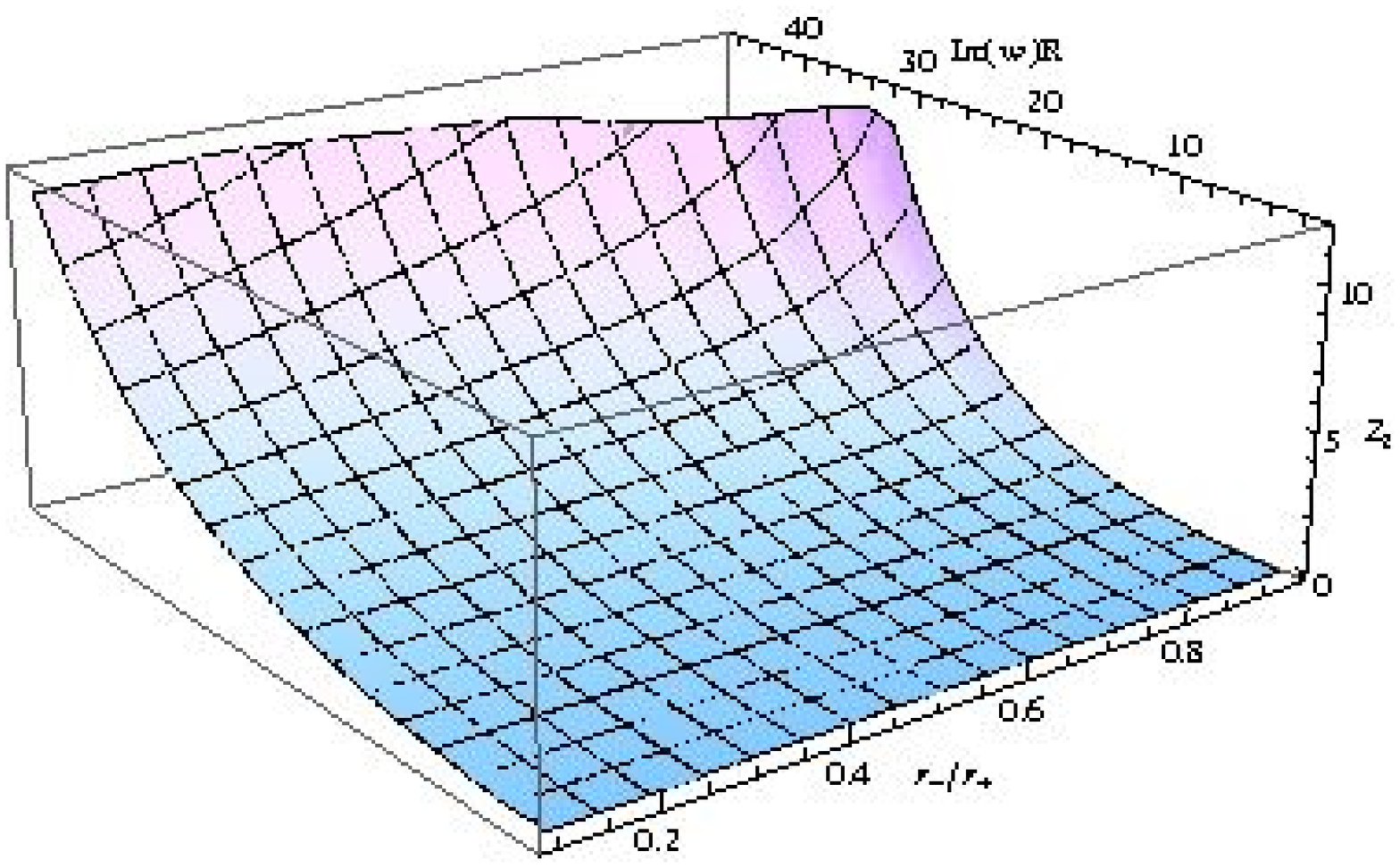}
\caption{$Z_i$ as a function of $r_-$ and $Im(\omega)$ for $D=6$, $r_+=10R$.}\label{D=6.r=10}
\end{figure}
\begin{figure}
\includegraphics[width=0.7 \linewidth]{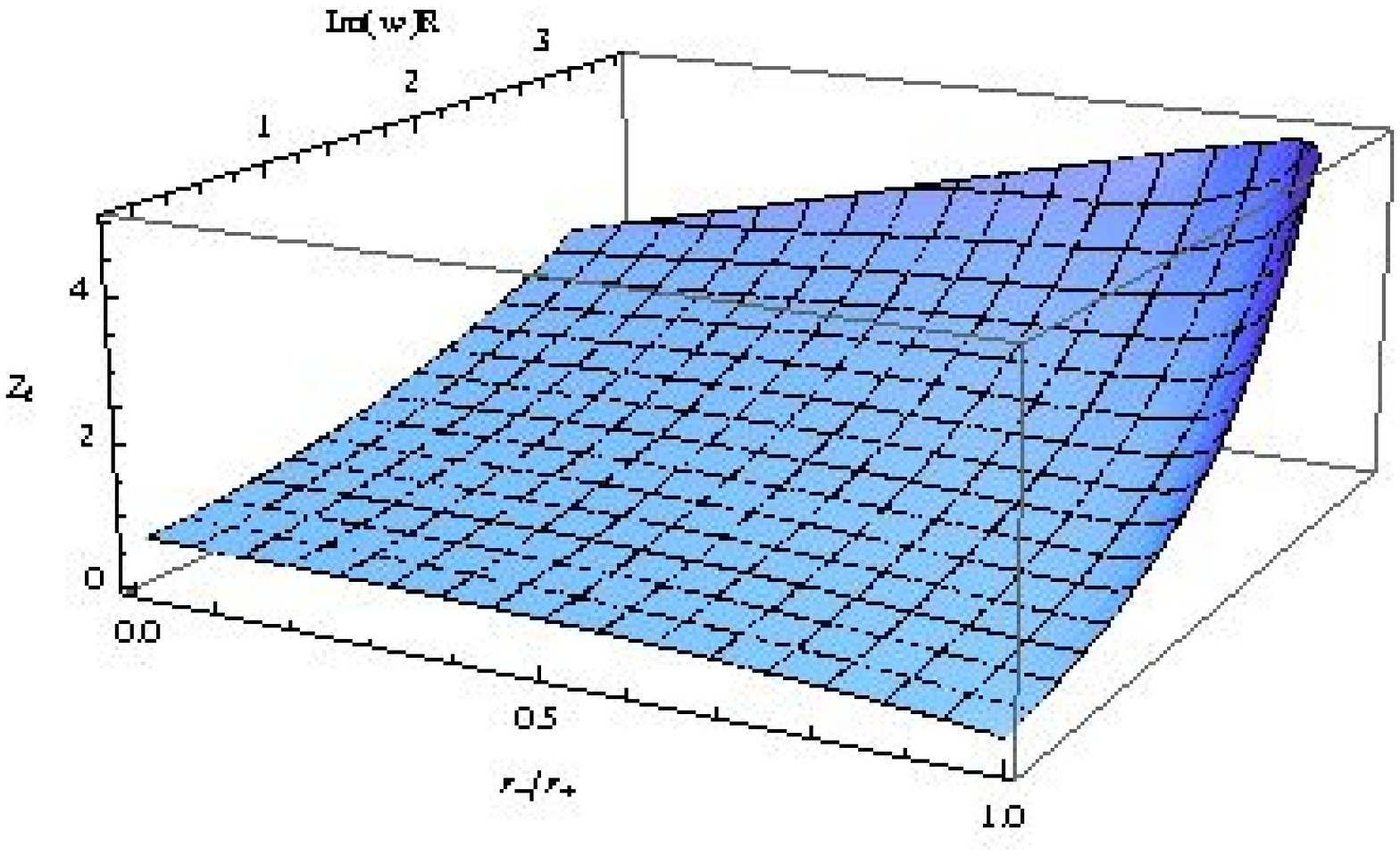}
\caption{$Z_i$ as a function of $r_-$ and $Im(\omega)$ for $D=7$, $r_+=R$.}\label{D=7.r=1}
\end{figure}
\begin{figure}
\includegraphics[width=0.7 \linewidth]{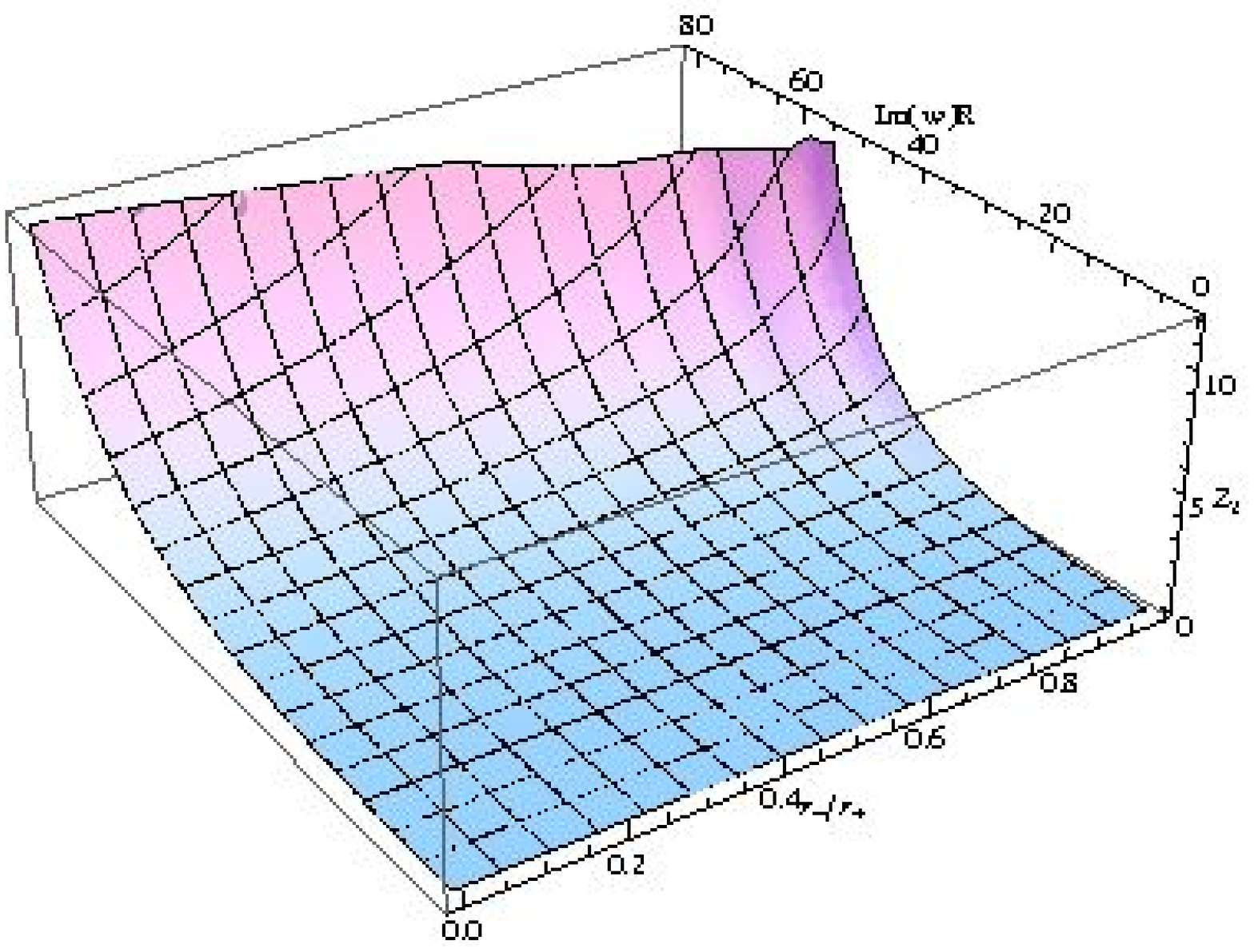}
\caption{$Z_i$ as a function of $r_-$ and $Im(\omega)$ for $D=7$, $r_+=10R$.}\label{D=7.r=10}
\end{figure}
\begin{figure}
\includegraphics[width=0.7 \linewidth]{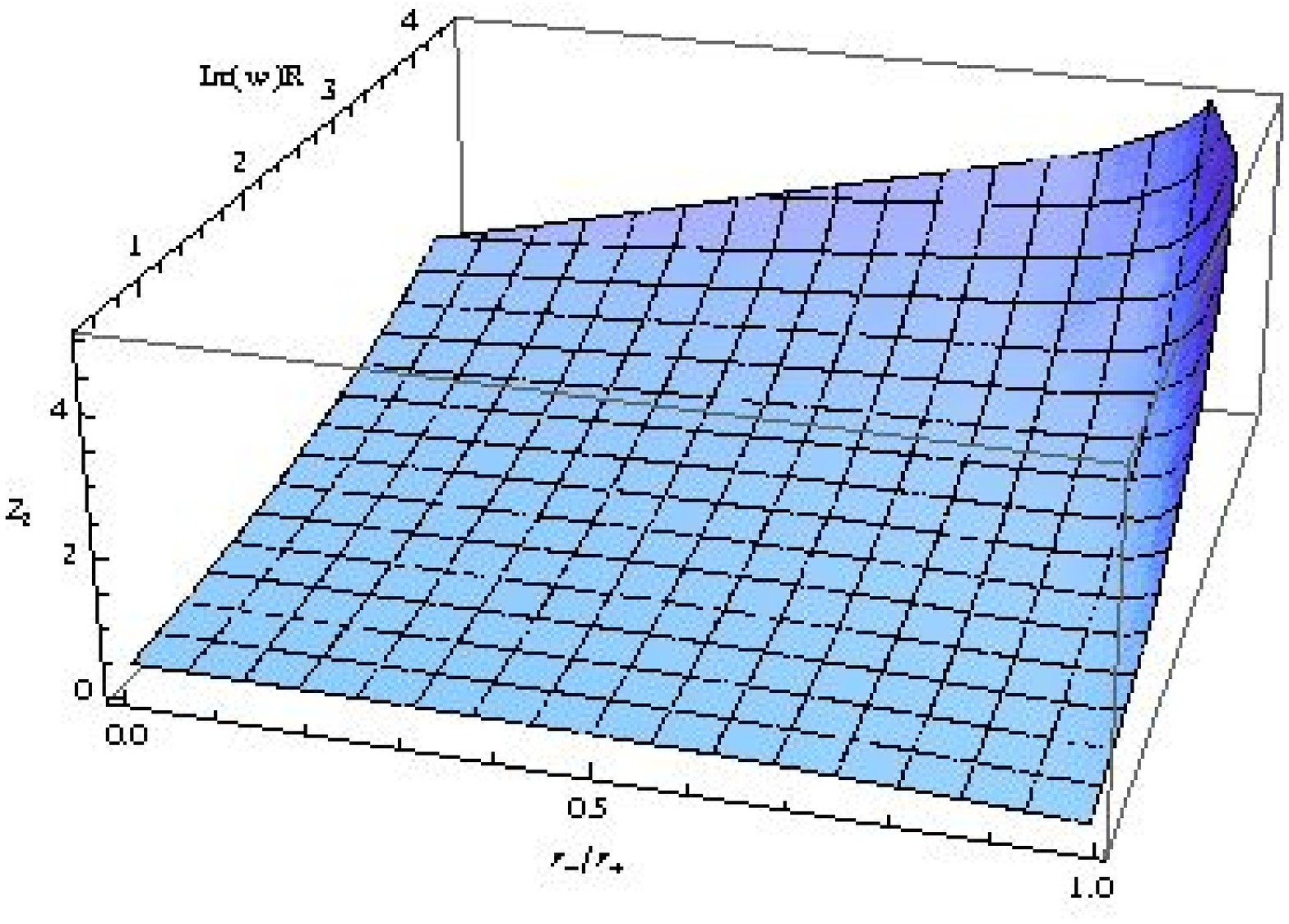}
\caption{$Z_i$ as a function of $r_-$ and $Im(\omega)$ for $D=8$, $r_+=1 R$.}\label{D=8.r=1}
\end{figure}
\begin{figure}
\includegraphics[width=0.7 \linewidth]{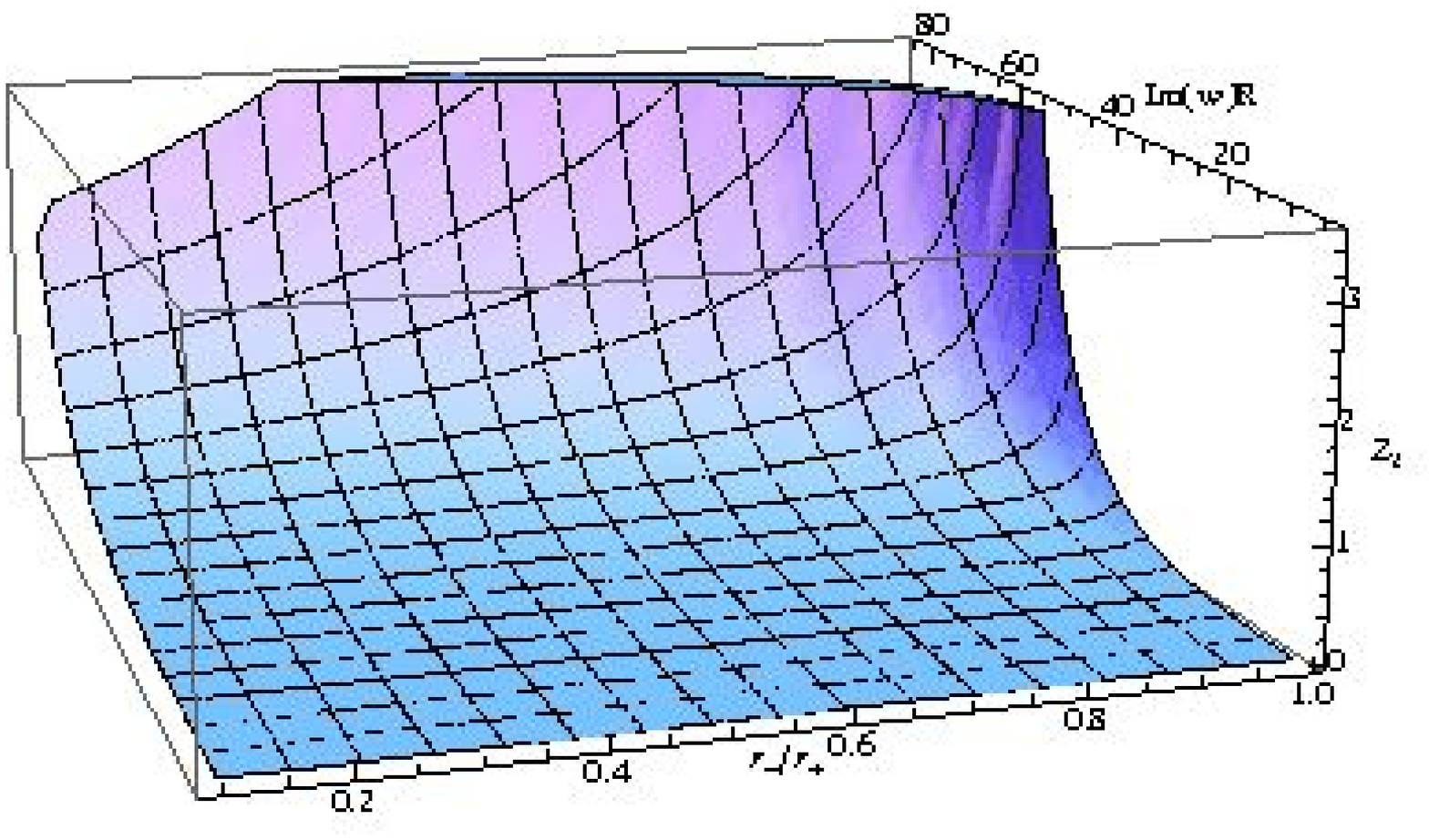}
\caption{$Z_i$ as a function of $r_-$ and $Im(\omega)$ for $D=8$, $r_+=10 R$.}\label{D=8.r=10}
\end{figure}
\begin{figure}
\includegraphics[width= 0.7 \linewidth]{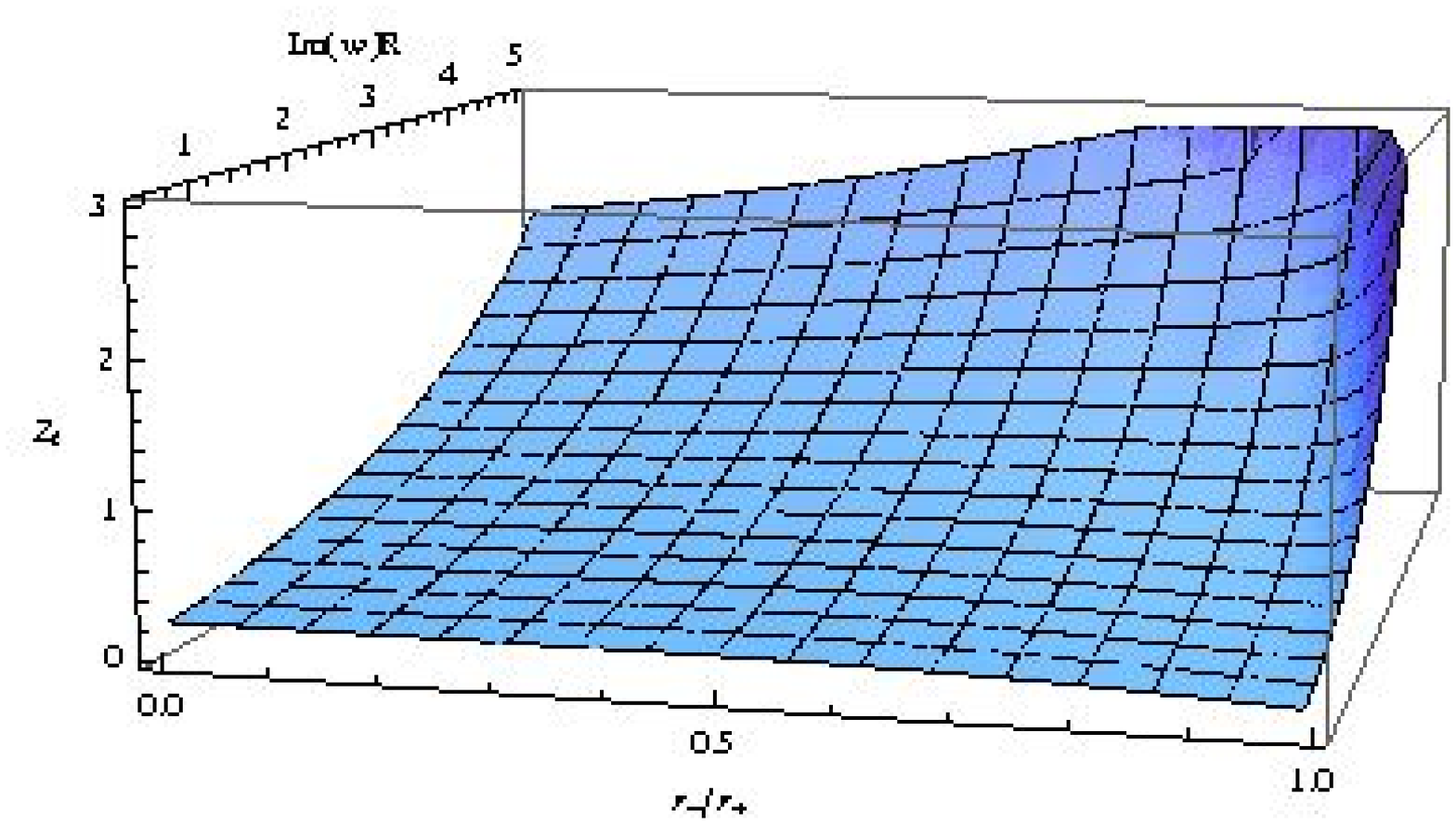}
\caption{$Z_i$ as a function of $r_-$ and $Im(\omega)$ for $D=9$, $r_+=1 R$.}\label{D=9.r=1}
\end{figure}
\begin{figure}
\includegraphics[width= 0.7 \linewidth]{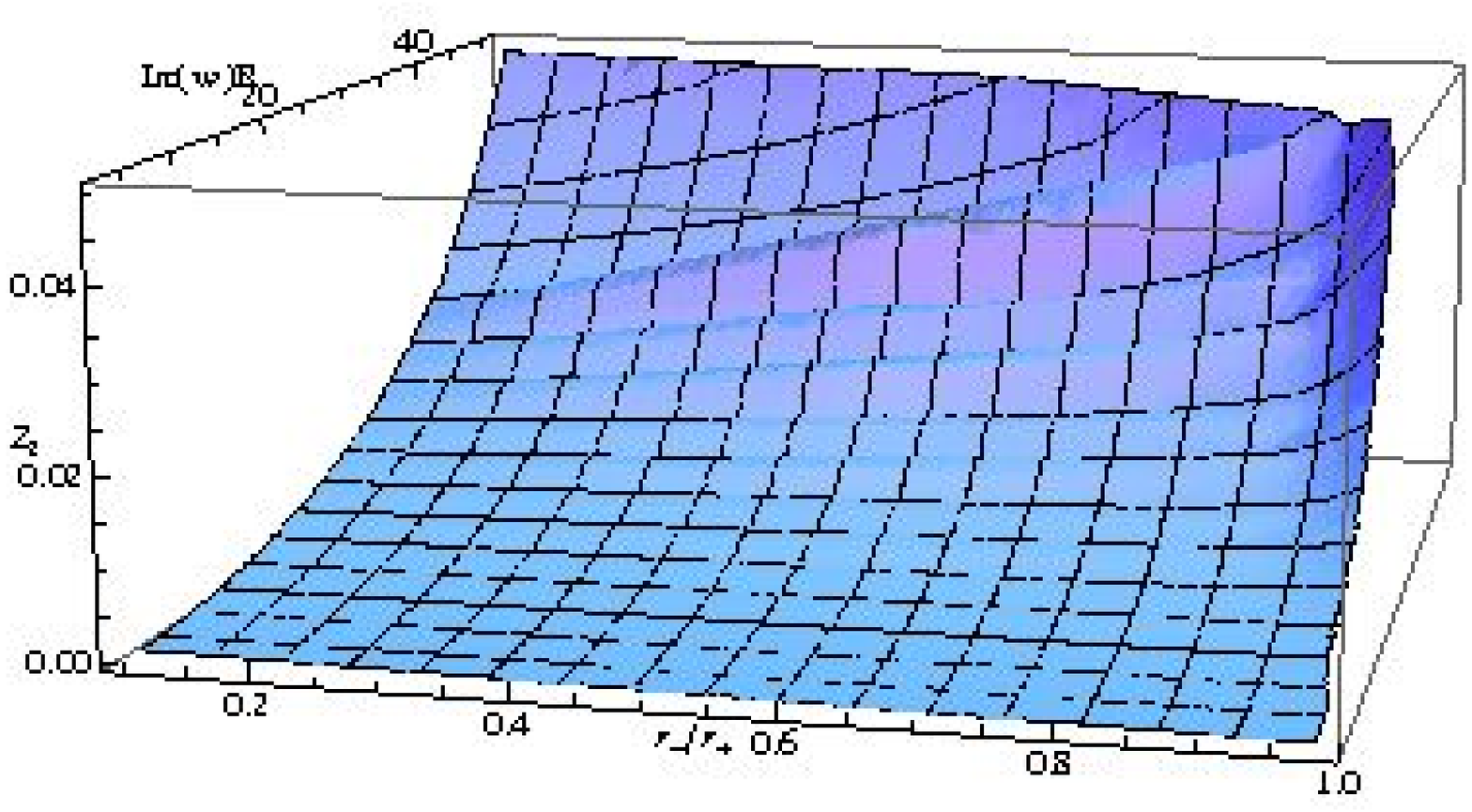}
\caption{$Z_i$ as a function of $r_-$ and $Im(\omega)$ for $D=9$, $r_+=10 R$.}\label{D=9.r=10}
\end{figure}
\begin{figure}
\includegraphics[width= 0.7 \linewidth]{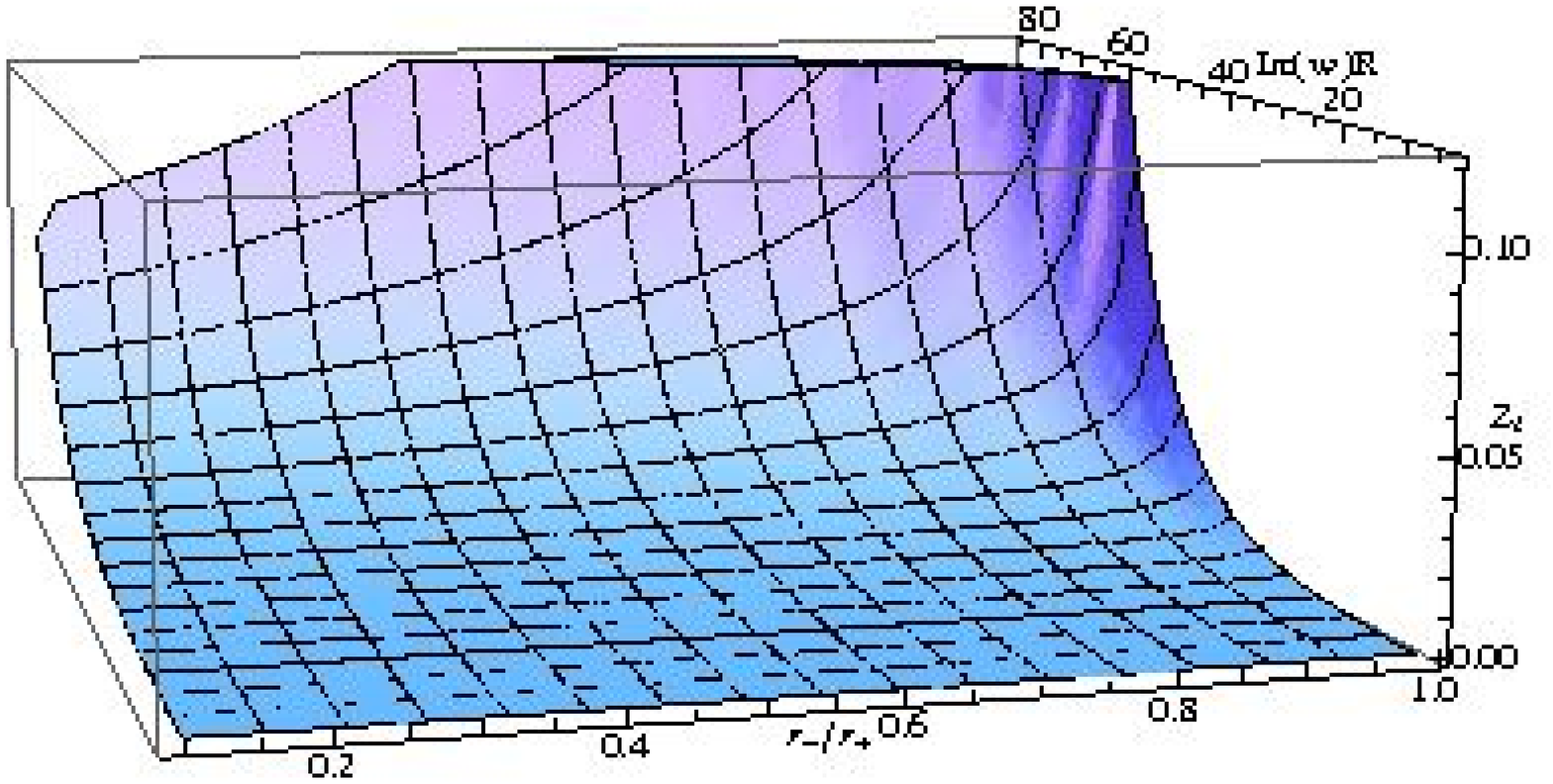}
\caption{$Z_i$ as a function of $r_-$ and $Im(\omega)$ for $D=10$, $r_+=10 R$; logarithmic plot}\label{D=10.r=10}
\end{figure}
\begin{figure}
\includegraphics[width= 0.7 \linewidth]{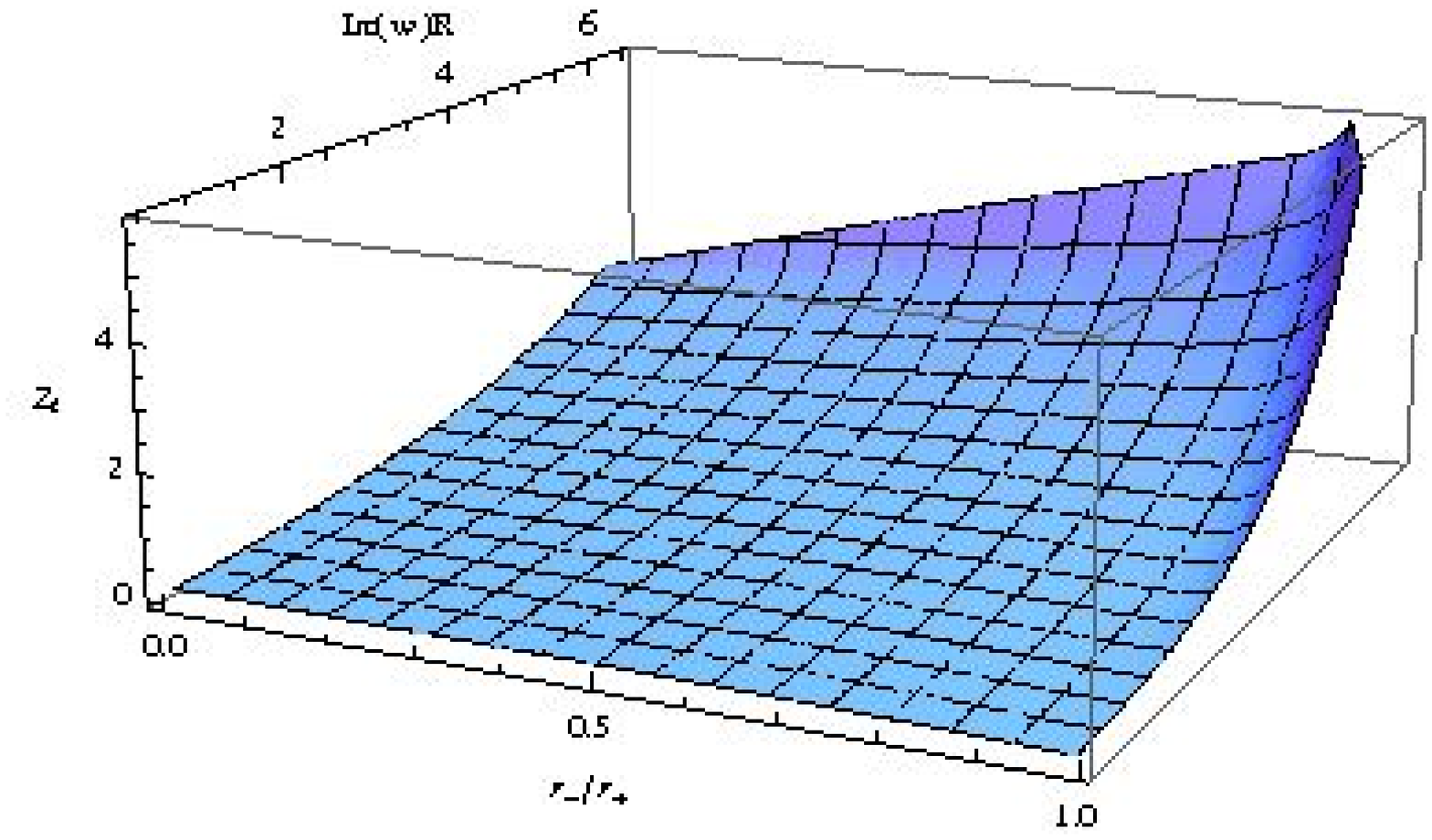}
\caption{$Z_i$ as a function of $r_-$ and $Im(\omega)$ for $D=10$, $r_+=1 R$; logarithmic plot}\label{D=10.r=1}
\end{figure}
\begin{figure}
\includegraphics[width= 0.7 \linewidth]{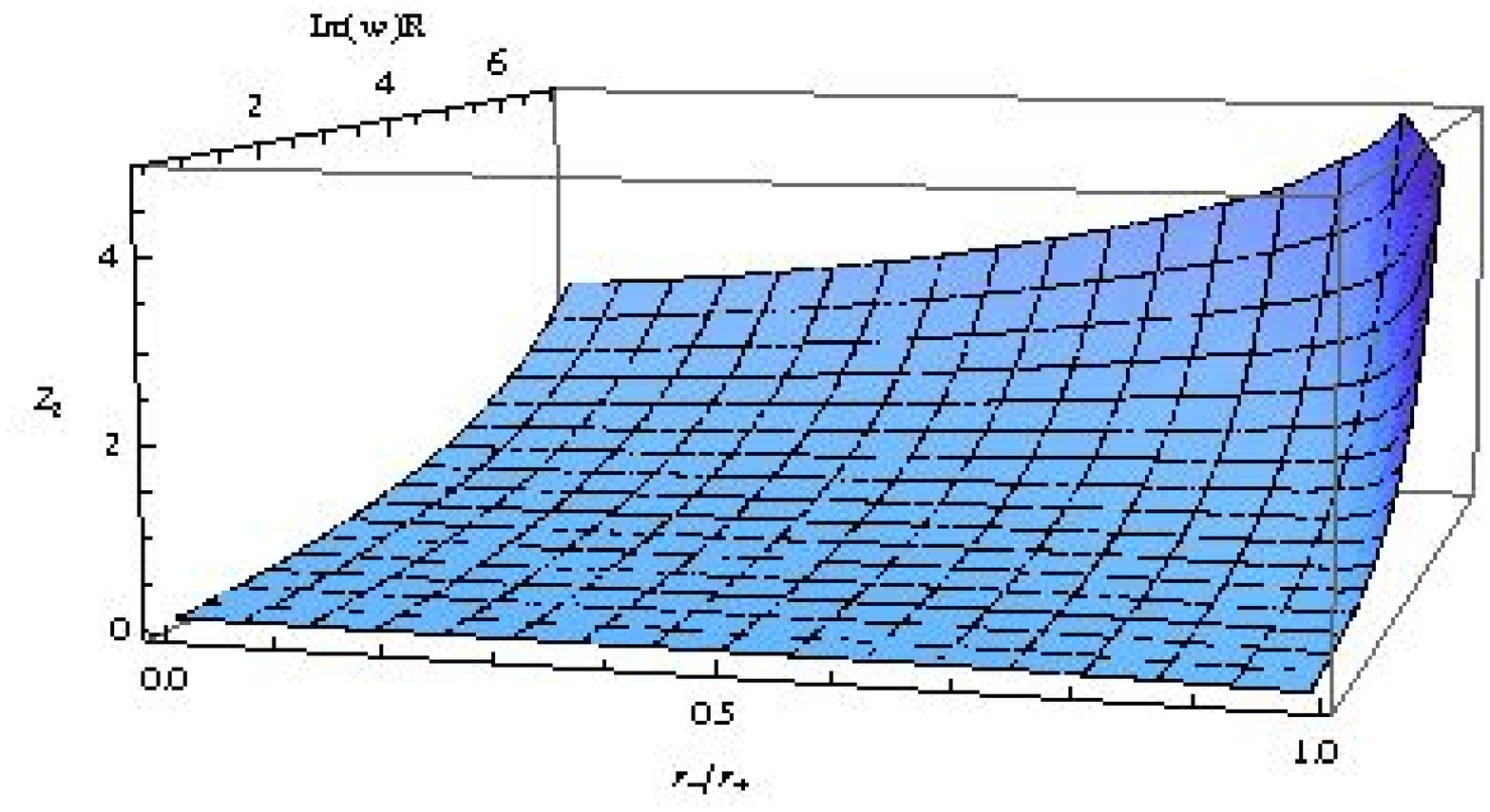}
\caption{$Z_i$ as a function of $r_-$ and $Im(\omega)$ for $D=11$, $r_+=1 R$}\label{D=11.r=1}
\end{figure}
\begin{figure}
\includegraphics[width= 0.7 \linewidth]{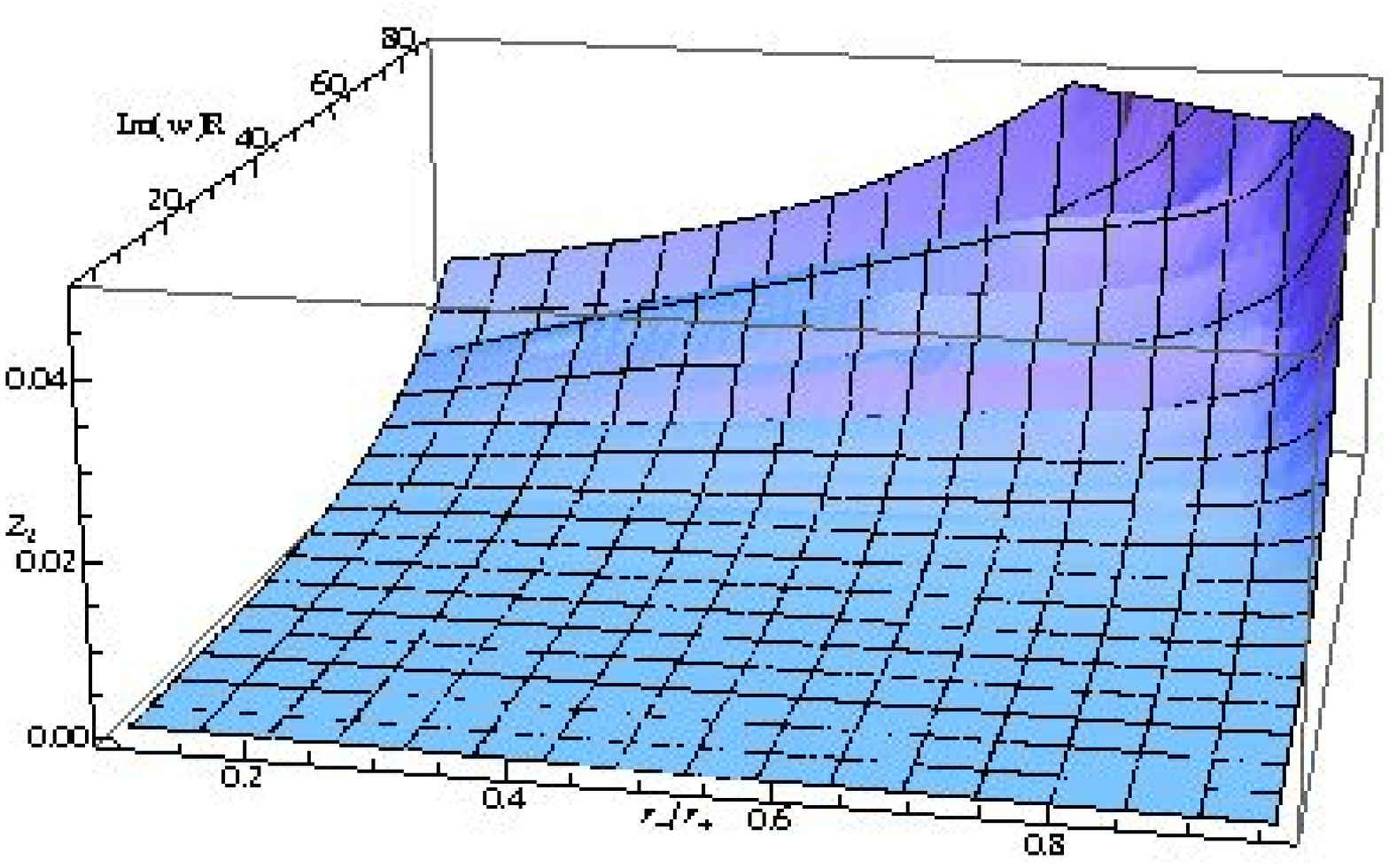}
\caption{$Z_i$ as a function of $r_-$ and $Im(\omega)$ for $D=11$, $r_+=10 R$}\label{D=11.r=10}
\end{figure}

\section{Stability analysis}

For testing the stability, we have to perform the two tasks: to check that the there is no unstable modes by the method described
in the previous section for the full range of the black hole parameters and, as a confirmation, to find the fundamental
(damped, when the system is stable) quasinormal modes.

\begin{figure}
\includegraphics[width= 0.7 \linewidth]{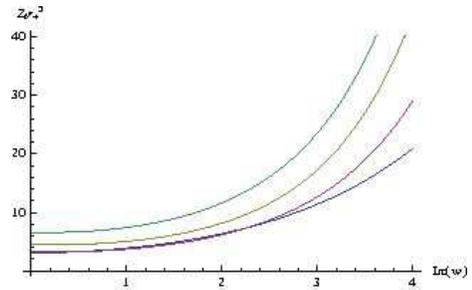}
\caption{$Z_i$ for small Schwarzschild-AdS black holes $D=5$; blue (top) for $r_+ = R$, red for $r_+ =R/2$,
yellow for $r_+ =R/4$, green (bottom) for $r_+ = R/8$.}\label{D=5.small}
\end{figure}

We shall distinguish here three regimes: large black holes $r_{+} \gg R$, intermediate black holes $r_{+} \sim R$
and small black holes $r_{+} \ll R$, where $R$ is the anti-de Sitter radius.
From the figures (\ref{D=5.r=1})-(\ref{D=11.r=10}), we can see that for the $D=5,6\ldots11$ black holes, $Z_i$ does not equal zero for any values of
$\omega$ limited by $Im(\omega)<\sqrt{-V_{min}},$. We have shown this here mainly for the two values of $r_{+}$: $r_{+} = 1 R$ and $r_{+} = 6 R$.
These are representative cases of large and intermediate AdS black holes. For small black holes, an example of $Z_i$ behavior
can be seen in fig. \ref{D=5.small} for $D=5$. There one can see that the smaller the size of the black hole, the larger $Z_i$, what guarantees no-instability at sufficiently small black hole size. Looking carefully at all range of parameters of $r_{+}$, $r_{-}$ and $\ell$, we have not found any zeros of $Z_i$. Therefore we conclude that $D=5,6\ldots11$ \emph{Reissner-Nordstr\"om-anti-de Sitter black holes are stable for any values of the black hole parameters}. Now, we shall check this by the search of the fundamental quasinormal modes, which, as it will be shown soon, all are damped.

The spectrum of frequencies of neutral asymptotically AdS black holes is qualitatively different from asymptotically
flat or de Sitter cases: the main striking feature of the spectrum is that almost all modes are proportional to the
radius of the black holes for large black holes $r_{+} \gg R$.
The exception is the fundamental mode of the scalar type of gravitational perturbations of SAdS black holes: its real part
approaches constant as $r_{+}$ goes to infinity, while the imaginary part is inverse proportional to
the radius of the black hole. Let us note that this property keeps also when AdS - black holes are charged.

Quasinormal modes of a particular case of large $D=5$ SAdS were considered in \cite{Friess:2006kw}. One can see in the table I
that we accurately reproduce their results. Indeed, the $Q = 0$, $r_{+} =6$ mode in our table I coincide with the fundamental mode
of  \cite{Friess:2006kw} in proper units (see table III).

\begin{figure}
\includegraphics[width= 0.7 \linewidth]{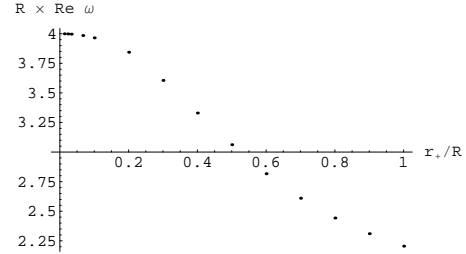}
\caption{$Re \omega_{n=0}$ for small $D=5$ Schwarzschild-AdS black holes approaches the limit $\omega =4$ as $r_{+} \rightarrow 0$ }\label{pureAdSplot}
\end{figure}

Another property of asymptotically AdS black holes is that their quasinormal modes approach the real normal modes of the pure
anti-de Sitter black holes, when the radius of the black hole goes to zero \cite{Konoplya:2002zu}. This was shown for the test
scalar field perturbations around $D=5$ and $6$ SAdS black holes in \cite{Konoplya:2002zu}. Here we have shown that for the scalar type of gravitational perturbations, quasinormal modes also reach their $D \geq 6$ pure anti-de Sitter values
\begin{equation}\label{pureAdS}
\omega_{n} R = 2 n +  D + \ell - 3, \quad D \geq 6 \quad pure \quad AdS
\end{equation}
The above formula for AdS space-time normal modes $\omega_n$ \cite{Natario:2004jd} is valid only for $D > 5$, while \emph{for $D=5$ }
\emph{the pure AdS spectrum is continuous}, i.e. all modes are normal modes of $D=5$ AdS space-time.
This happens because of the peculiar behavior of the effective potential at spatial infinity (which is zero in the tortoise coordinate):
the effective potential has infinite negative pitch near $r^{*} = 0$ (see Fig. \ref{D=5.r=10potential}), so that $\Psi \rightarrow 0$ for
all $\omega$ \cite{Natario:2004jd}. Notice that in spite of the infinite negative pitch, the area covered by the pitch is finite, so that the instability is not guaranteed \emph{a priori} even for this case.  The natural question arises: if the limit of small
black holes has the same meaning for $D=5$ as it has for $D \geq 6$?
On fig. \ref{pureAdSplot} one can see, that in the limit of small black holes, the $D=5$ QNMs approach the same limit as $D \geq 6$ modes do.

It is not remarkable that for Reissner-Nordstr\"om-AdS black holes, QNMs approach the same pure AdS values in the limit $r_{+} \rightarrow 0$ (\ref{pureAdS}), because one cannot assume $r_{+} = 0$, without taking $Q=0$.
Thus, the fundamental quasinormal modes of the scalar type of gravitational perturbations of D-dimensional RNAdS black holes obey
\begin{equation}
\omega_{n} R \rightarrow 2 n + D + \ell - 3, \quad r_{+} \rightarrow 0, \quad D \geq 5
\end{equation}

Let us note that once we proved here the stability of the D-dimensional Schwarzschild-AdS black holes, the stability of the
Reissner-Nordstr\"om-AdS black holes can be intuitively understood from the behavior of the effective potentials (figures for scalar type in \cite{ishibashi_kodama2}, \cite{ishibashi_kodama}) at least for $D \geq 6$: the presence of the charge $Q$ increases slightly the negative depth of the potential gap. Apparently the negative gap is not deep enough to allow bound states with negative "energy". The behavior of the effective potential  for $D=5$ is quite different (\ref{D=5.small}), yet, as we have shown, this does not lead to instability as well.

\begin{table}
\caption{Fundamental ($n=0$) quasinormal modes of $D=5$ Reissner-Nordstr\"om-AdS black holes.}
\begin{tabular}{|c|c|c|}
  \hline
  $r_{-}/r_{+}$ & $D=5$, $r_{+} =1$ & $D=5$, $r_{+} =6$ \\
  \hline
  0 & 2.20477 - 0.58137 i & 1.65171 - 0.13759 i \\
  0.1 & 2.19646 - 0.57776 i & 1.65143 - 0.13623 i \\
  0.2 & 2.17255 - 0.56754 i & 1.65058 - 0.13214 i \\
  0.3 & 2.13571 - 0.55236 i & 1.64914 - 0.12544 i \\
  0.4 & 2.08934 - 0.53445 i & 1.64713 - 0.11639 i \\
  0.5 & 2.03698 - 0.51637 i & 1.64455 - 0.10549 i \\
  0.6 & 1.98248 - 0.50074 i & 1.64150 - 0.09350 i \\
  0.7 & 1.93093 - 0.48956 i & 1.63800 - 0.08131 i \\
  0.8 & 1.88885 - 0.48091 i & 1.63404 - 0.06989 i \\
  0.9 & 1.85619 - 0.46802 i & 1.62944 - 0.06059 i \\
  0.99 & 1.89919- 0.24619 i & 1.62648 - 0.05644 i \\
  \hline
\end{tabular}
\end{table}

\begin{table}
\caption{Fundamental ($n=0$) quasinormal modes of $D=6, 7$ Reissner-Nordstr\"om-AdS black holes.}
\begin{tabular}{|c|c|c|}
  \hline
  $r_{-}/r_{+}$ & $D=6$, $r_{+} = 6$ & $D=7$, $r_{+} = 6$  \\
  \hline
  0 & 1.59784 - 0.14933 i & 1.56442 - 0.15517 i  \\
  0.1 & 1.59781 - 0.14918 i & 1.56442 - 0.15515 i  \\
  0.2 & 1.59761 - 0.14814 i & 1.56438 - 0.15492 i  \\
  0.3 & 1.59708 - 0.14533 i & 1.56420 - 0.15392 i  \\
  0.4 & 1.59603 - 0.13994 i & 1.56372 - 0.15123 i  \\
  0.5 & 1.59430 - 0.13138 i & 1.56270 - 0.14567 i  \\
  0.6 & 1.59175 - 0.11954 i & 1.56086 - 0.13607 i  \\
  0.7 & 1.58832 - 0.10503 i & 1.55790 - 0.12180 i  \\
  0.8 & 1.58395 - 0.08923 i & 1.55359 - 0.10364 i  \\
  0.9 & 1.57829 - 0.07427 i & 1.54753 - 0.08409 i  \\
  0.99 & 1.57267 - 0.06664 i & 1.53968  - 0.07200 i \\
  \hline
\end{tabular}
\end{table}

\section{Discussions}
In this paper, by the numerical search of quasinormal modes, we have shown that Reissner-Nordstr\"om-anti-de Sitter black holes are gravitationally stable in $D=5,6\ldots11$ space-time dimensions. Before, the stability of asymptotically anti-de Sitter black holes was established only for $D=4$ Reissner-Nordstr\"om-anti-de-Sitter black holes analytically \cite{ishibashi_kodama}. Stability for $D=5-11$ found here and for $D=4$ found by Ishibashi and Kodama \cite{ishibashi_kodama} does not contradict to the observed instability for $D=4, 5$ RNAdS black holes by Gubser and Mitra in \cite{Gubser:2000mm}, \cite{Gubser:2000ec}, because the latter instability is induced by a tachyonic field coupled to the system in  the $\mathcal{N}=8$ gauged supergravity. Thus, although metrics for the black hole in both cases are the same RNAdS metric, they are exact solution of different field equations, and the dynamic of perturbed equations is certainly different.

The observed here stability of RNAdS black holes is interesting also, because we know that small AdS black holes (\emph{within the ordinary Einstein-Maxwell theory, considered here}) are thermodynamically unstable and may exert the Hawking-Page transition.
It would be natural to expect that this thermodynamic transition will be accompanied by a gravitational instability.
Yet, as we have shown here, this does not take place, so that if the correlation between thermodynamic and gravitational instabilities exists, it is
more subtle, than one could naively expect for complex gravitational systems.

An important question, which was beyond the scope of our work, is the stability of extremally charged RNAdS black holes.
Our closest aim is to give a detailed data on quasinormal modes of other types of gravitational perturbations (vector and tensor), and to find higher overtones of the spectrum \cite{Siopsis-us}. We believe it would be interesting to investigate stability of charged asymptotically AdS black holes
in the Gauss-Bonnet theory, where already there is instability, stipulated by Gauss-Bonnet terms, at higher multipoles.

\begin{acknowledgments}
We would like to acknowledge Andrei Starinets for very valuable discussions.  R.K. acknowledges support of the Japan Society for the Promotion of Science (JSPS). A. Z. was supported by \emph{Funda\c{c}\~ao de Amparo \`a Pesquisa do Estado de S\~ao Paulo (FAPESP)}, Brazil.
\end{acknowledgments}


\end{document}